\documentclass{article}
\usepackage[T1]{fontenc}
\usepackage[utf8]{inputenc}
\usepackage[numbers,sort&compress]{natbib}
\usepackage{url}
\usepackage{hyperref}
\usepackage[version=4]{mhchem}

\usepackage[a4paper,margin=2.5cm]{geometry}
\usepackage{graphicx}

\begin{document}

\markboth{Languin-Catto\"en \& Bussi}{RNA Dynamics and Interactions}

\title{RNA Dynamics and Interactions Revealed through Atomistic Simulations}

\author{Olivier Languin-Catto\"en and Giovanni Bussi\footnote{bussi@sissa.it}\\
Scuola Internazionale Superiore di Studi Avanzati, SISSA, Trieste, Italy, 34136}
\maketitle
\thanks{This manuscript has been published in \textit{Annual Review of Physical Chemistry}, Volume 77 (2026), DOI: 10.1146/annurev-physchem-082624-013453. The present document is the author-accepted manuscript. Posted with permission from the \textit{Annual Review of Physical Chemistry}, Volume 77; copyright 2026 the author(s), https://www.annualreviews.org.}

\begin{abstract}
RNA function is deeply intertwined with its conformational dynamics. In this review, we survey recent advances in the use of atomistic molecular dynamics simulations to characterize RNA dynamics in diverse contexts, including isolated molecules and complexes with ions, small molecules, or proteins. We highlight how enhanced sampling techniques and integrative approaches can improve both the precision and accuracy of the resulting structural ensembles. Finally, we examine the emerging role of artificial intelligence in accelerating progress in RNA modeling and simulation.
\end{abstract}


\section{INTRODUCTION}

Ribonucleic acid (RNA)
molecules play a fundamental role in life sciences,
transiently storing genetic material in the form of messenger RNAs (mRNAs), mediating protein synthesis
as transfer and ribosomal RNAs (tRNA and rRNA),
and regulating gene expression in complex organisms \cite{morris_rise_2014}. Due to their ubiquitous role and to the difficulties in targeting proteins, RNA molecules are considered promising drug targets \cite{childs-disney_targeting_2022}.
Furthermore, RNA vectors have revolutionized medicine, impacting the fight against global pandemics
and offering a range of other potential applications \cite{sahin_mrna-based_2014}.
RNA structure is hierarchical \cite{tinoco_how_1999} and classified as primary (the nucleotide sequence), secondary (the list of canonical helices, forming Watson-Crick pairs)
and tertiary (the arrangement of these helices
in three-dimensional space, as well as the additional non-canonical contacts).
In the traditional view, the function of nucleic acids only depends on their primary structure,
which encodes information.
The modern interpretation is however that secondary and tertiary structures are also connected to RNA function.
The primary structure is fundamental when considering interactions between RNA and other nucleic acids, often dominated by the specific recognition of Watson-Crick base pairs. Similarly, thanks to the adapter role played by tRNAs, primary structure of the coding portions of mRNAs is used to store protein sequence information.
Secondary structure determines which parts of the primary structure are available for pairing, that is, which parts of the  message
are readable. Tertiary structure is instead crucial whenever considering the interaction between RNA molecules and ions, proteins, or other molecules~\cite{butcher_molecular_2011}.

RNA molecules can be highly dynamic. Here, ``dynamics'' refers to the coexistence of multiple conformations in equilibrium, and should not be confused with ``kinetics'', which concerns the time scales of transitions between conformations. These alternative conformations may lie within a narrow free-energy range and exhibit comparable populations.
This can happen at the secondary  \cite{bose_causes_2024} or tertiary \cite{ken_rna_2023} structure level.
RNA dynamics can be very important when binding other molecules, because intermolecular interactions
can result in conformational selection or induced-fit processes. However, individuating multiple concurrent structures in experiments is difficult.

\begin{figure}
    \centering
    \includegraphics[width=1\linewidth]{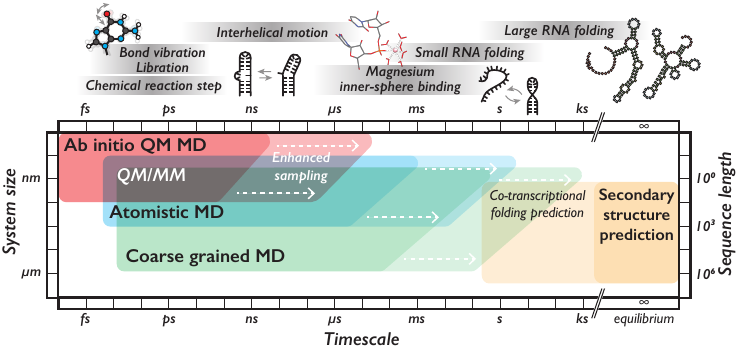}
    \caption{Typical time scales of RNA dynamic processes and corresponding simulation methods.
Molecular dynamics (MD) simulations based on quantum mechanics (QM), hybrid quantum mechanics/molecular mechanics (QM/MM), classical atomistic force fields, and coarse-grained (CG) models can be used to study RNA dynamics across a broad range of time scales and system sizes. Enhanced sampling techniques can extend the effective time scales accessible to molecular simulations. In parallel, RNA secondary-structure prediction methods can be used to model equilibrium ensembles or to mimic the long-time-scale behavior of RNA dynamics.
    }
    \label{fig:timescales}
\end{figure}

Atomistic molecular dynamics (MD) simulations explicitly model the system of interest at atomic resolution using physics-based principles. The employed energy models---force fields---are parameterized using a mixture of experimental data and quantum-chemistry calculations \cite{frohlking_toward_2020}.
Biomolecular simulations
provide detailed atomistic pictures of conformational ensembles and have been interpreted as computational microscopes able to complement many experimental methodologies \cite{lee_discovery_2009}.
The time and length scales accessible to various simulation methods are schematized in
\textbf{Figure~\ref{fig:timescales}}.
At the smallest scales we have quantum mechanical (QM) calculations, which explicitly
describe the chemistry of the investigated systems and can be used to characterize chemical reactions,
such as those catalyzed by ribozymes or those occurring when chemical probes covalently bind to RNA.
They also include polarization effects, that are important for accurately describing
the interaction between RNA molecules and charged groups.
By contrast, classical MD simulations of RNA systems are typically based on force fields that neglect both charge transfer and polarization, with only a few studies employing polarizable models.
At larger scales we have coarse-grained (CG) models, which can be effectively
used to simulate large conformational transitions. However, they typically require ad hoc
parameterization to be usable in different contexts. For instance, CG models
that can accurately characterize non-Watson-Crick base pairings are still rare.
At even higher time and length scales we can find thermodynamic models for RNA secondary structure.
It is worth mentioning that these methods can be combined
in a multi-scale fashion, as is done, for instance, in quantum-mechanics/molecular-mechanics (QM/MM) simulations.

\begin{figure}
    \centering
    \includegraphics[width=\linewidth,trim=0 10 0 0,clip]{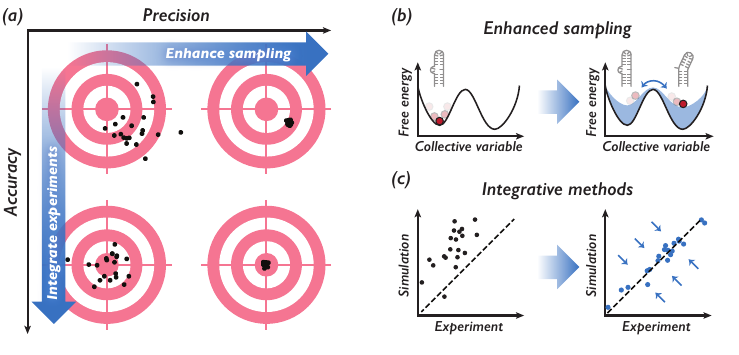}
    \caption{
    Molecular simulations can suffer from both limited accuracy and limited precision \textit{(a)}. Precision can be improved using enhanced sampling methods, which reduce free-energy barriers and enable better exploration of conformational space \textit{(b)}. Accuracy can be improved by integrating experimental data, either by refining force-field parameters or reweighting simulated ensembles to better match observations \textit{(c)}.
    }
    \label{fig:introduction}
\end{figure}

Molecular simulations might suffer from both precision and accuracy issues
(\textbf{Figure~\ref{fig:introduction}}).
Precision is the capability of a simulation method to produce consistent results independently
of the initial conditions, and hence depends on the time scale on which it is evaluated and on the fraction
of the conformational space that can be explored.
Precision can be improved by running longer simulations, or by taking advantage of enhanced sampling methods.
Simulations on different scales have different precision issues. QM simulations usually
only cover a small fraction of the conformational space compared to classical MD simulations and CG models. For sufficiently small systems,
the latter two can sample ergodically the conformational space, so that the observed
trajectory statistically converges to a well-defined thermodynamic ensemble.
To achieve this level of extensive sampling, it is often necessary to use enhanced sampling methods, which typically sacrifice kinetics in favor of faster convergence of equilibrium properties.
Thermodynamic models can achieve infinite precision by virtually enumerating
all possible secondary structures for systems of length up to several thousand nucleotides.

Independently of precision, molecular simulations might suffer from poor accuracy, i.e.~their capability
to reproduce experimental data.
In principle, QM calculations closely represent the laws of the physical world.
However, they can rarely be performed with sufficient accuracy on molecules of relevant size.
Classical MD simulations, CG models, and thermodynamic models are typically parameterized
using an increasing fraction of experimental information.
Molecular simulations should be validated against experiments.
When possible, experimental information should be used to improve their accuracy.

In this review, we provide an overview of recent works addressing RNA structural dynamics using classical atomistic molecular
simulations. To complement ours, we refer the interested reader to other reviews: Reference~\citenum{sponer_rna_2018} exhaustively covers methods and applications up to 2018,
whereas Reference~\citenum{muscat_power_2024} provides a recent perspective, with a different selection of applications compared to this review.
More specific reviews will be mentioned in the following, when appropriate.

\section{METHODOLOGICAL BACKGROUND}
This section provides general background on MD simulations of RNA systems. MD simulations involve integrating the equations of motion for a model in which atoms are typically treated as point particles. General references on MD simulations are available in standard textbooks (see, e.g., Reference~\citenum{tuckerman_statistical_2023}). Below, we discuss in more detail the specific considerations relevant to RNA systems.

\subsection{Currently used force fields for RNA system}

In classical MD simulations, interactions are computed using empirical force fields
parameterized from a blend of experimental data and quantum-chemistry calculations 
\cite{frohlking_toward_2020}.
Different families of RNA force fields have been introduced.
Exhaustive lists can be found in other reviews \cite{sponer_rna_2018,muscat_power_2024,liebl_development_2023}.
Here, we instead focus on some of the most recent developments.

The most extensively tested force field from the AMBER family is $\chi$OL3 \cite{zgarbova_refinement_2011}.
Recent empirical corrections have adjusted the strength of specific hydrogen bonds
\cite{frohlking_automatic_2022}, modified Lennard-Jones interactions \cite{mlynsky_simple_2023,love_van_2024},
or applied alternative charge-derivation strategies \cite{janecek_w-resp_2021}.
A more extensive re-parameterization has been proposed in the D. E. Shaw lab
\cite{tucker_development_2022}.
The most recent and reliable force field from the CHARMM family is CHARMM36 \cite{denning_impact_2011}.
These additive force fields are employed in most of the simulations discussed in this review.
It is also important to note the possibility of using polarizable force fields
\cite{zhang_amoeba_2018,lemkul_polarizable_2018}.
Commonly used water models include TIP3P, TIP4P, SPC/E, OPC, TIP4P-D
(see References \citenum{kuhrova_sensitivity_2023} and \citenum{sarthak_benchmarking_2023} for
recent tests of these models in combination with RNA molecules).

The validity of force fields
should be tested against experimental observations.
Recent papers provide extensive tests on benchmark systems
and can be a useful starting point to assess whether a specific system
is described well by one of the available force fields
\cite{sarthak_benchmarking_2023,winkler_benchmarking_2023,mlynsky_can_2025,lemmens_kink-turn_2025}.

\subsection{Integrative methods}

Imperfect force fields may lead to discrepancies between simulation results and experimental observations.
A possible strategy to alleviate this problem is to integrate simulations and experiments.
This can be done following different philosophies.
In ensemble refinement methods, conformational ensembles generated by MD simulations are corrected to enforce agreement with experimental data.
These corrections can also be applied on-the-fly, aiming to maximize the overlap between sampled and experimental conformations.
Alternatively, force-field parameters can be directly fine-tuned to reproduce experimental observables.
For a more in-depth discussion of the pros and cons of these approaches, we refer the reader to a recent review focused on RNA systems \cite{bernetti_integrating_2023}.

\subsection{Enhanced sampling methods}

Modern hardware can typically support simulation times on the order of tens of microseconds for a typical biomolecular system ($10^5$--$10^6$ atoms including solvent), reaching up to milliseconds of cumulative simulation time in the most exhaustive studies. However, the time scales required for conformational sampling in RNA can extend to milliseconds for processes such as divalent cation binding and unbinding, and up to seconds or longer for complex secondary structure rearrangements (see \textbf{Figure~\ref{fig:timescales}}). Enhanced sampling methods are often employed to alleviate this issue.
Broadly speaking, some methods accelerate the entire system, for example by increasing the temperature or scaling down interatomic interactions.
Other methods target specific transitions across free-energy barriers, offering greater efficiency but requiring prior knowledge of relevant collective variables.
For a comprehensive introduction to these techniques, we refer the reader to a recent in-depth review \cite{henin_enhanced_2022}.
For a more RNA-oriented view, please turn to Reference~\citenum{mlynsky_exploring_2018}.

\section{SIMULATIONS OF ISOLATED RNA MOLECULES}

In this Section we discuss recent studies using MD simulations to characterize conformational sampling of RNA systems.

\begin{figure}
    \centering
    \includegraphics[width=\linewidth,trim=0 0 0 5,clip]{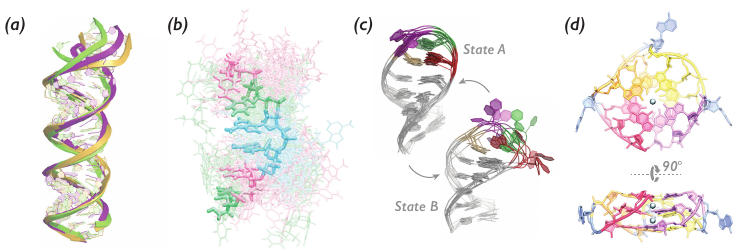}
    \caption{Examples of RNA systems investigated with MD simulations. \textit{(a)} Canonical A-form duplex simulated in various salt conditions, adapted from He et al.~\cite{he_structural_2021}. \textit{(b)} RNA hexamer (UCAAUC), and its highly diverse conformational ensemble, trajectory from Fröhlking et al.~\cite{frohlking_simultaneous_2023}. \textit{(c)} Hairpin with a UUCG tetraloop showing two states with distinct plasticity, adapted from Bottaro et al.~\cite{bottaro_integrating_2020}. \textit{(d)} Top and side views of an RNA G-quadruplex stabilized with \ce{K+}, coordinates from Pokorná et al.~\cite{pokorna_rna_2025-1}.}
    \label{fig:structures}
\end{figure}

\subsection{Canonical duplexes}

Canonical B-form helices have long served as benchmarks for deoxyribonucleic acid (DNA) simulations due to their ubiquitous
recurrence and the non-trivial dynamics they exhibit \cite{liebl_accurate_2021,lopez-guell_correlated_2023}. In contrast, RNA duplexes adopt an A-form helical structure (\textbf{Figure~\ref{fig:structures}a}), which is more rigid but equally important as a minimal model for testing RNA force fields. As such, simulations of RNA A-form helices represent a fundamental benchmark for MD simulations.

Detailed analyses of helical parameters have been used in several studies to assess and refine RNA-specific force fields
\cite{janecek_w-resp_2021,chen_rna-specific_2022,
liang_generalized_2025,kuhrova_sensitivity_2023}. Comparisons with experimental data, such as electron paramagnetic resonance measurements, have provided additional validation of simulated conformations \cite{gauger_structure_2024}. The sequence dependence of correlated motions in RNA duplexes has also been extensively characterized, for instance through simulations of a library of duplexes containing 136 unique tetramer sequences \cite{battistini_sequence-dependent_2023}. Furthermore, integrative approaches combining simulations with wide-angle X-ray scattering (WAXS) data have been applied to generate accurate structural ensembles of both RNA and DNA duplexes \cite{he_structural_2021}.

RNA--DNA hybrids are also important in biological processes such as transcription and gene editing.
Simulating these hybrids requires a careful combination of compatible RNA and DNA force fields,
which is complicated by the fact that these force fields have traditionally been developed and tested separately.
A recent systematic benchmark covering AMBER variants, CHARMM, and polarizable force fields concluded that none of them were able to reproduce the experimental structures of RNA--DNA hybrids \cite{knappeova_comprehensive_2024}.

\subsection{Unstructured RNA oligomers}

RNA tetramers and hexamers (\textbf{Figure~\ref{fig:structures}b}) have become a standard reference for testing both RNA force fields and enhanced sampling methods, largely thanks to a series of high-quality experimental studies from the Turner group.
Although these short sequences are not  biologically functional, their small sizes allow for highly detailed nuclear magnetic resonance (NMR) studies, in which all relevant peaks can be assigned. This facilitates the use of both positive and negative evidence (including missing peaks) to validate conformational ensembles.
Such comprehensive experimental datasets have been instrumental in providing a ground truth necessary to identify force-field deficiencies and guide improvements in RNA simulations \cite{
janecek_w-resp_2021,zhao_nuclear_2022,bergonzo_conformational_2022,he_refining_2022,frohlking_automatic_2022,chen_rna-specific_2022,li_base-specific_2022,mlynsky_simple_2023,winkler_benchmarking_2023,
liang_generalized_2025}.
Oligomers have also served as prototypical models for testing integrative approaches that combine simulations with experimental data \cite{frohlking_simultaneous_2023,gilardoni_boosting_2024}.
Importantly, these systems are typically dominated by base stacking interactions and lack intra-molecular hydrogen bonds. As such, they do not fully represent structured RNA motifs, but they are ideal for probing sequence-dependent stacking dynamics. For example, the tetramer A$_4$ (a single-stranded AAAA sequence) exhibits strong stacking, whereas sequences like C$_4$ or U$_4$ are considerably more flexible and disordered \cite{bottaro_conformational_2018}.

Beyond these minimal models, longer unstructured single-stranded RNAs have also been studied. A fragment-based approach was proposed in which simulations of tetramers were used to build a library for assembling conformational ensembles of longer RNAs \cite{pietrek_hierarchical_2024}. Alternatively, direct simulations of 30-nucleotide single-stranded RNAs with different sequence compositions have been performed using experimental restraints from small-angle X-ray scattering (SAXS) and reweighting from
Förster resonance energy transfer data to construct conformational ensembles consistent with observations \cite{wang_sequence-dependent_2024}.

\subsection{Hairpin loops}
\label{sec:hairpin-loops}

Hairpins are ubiquitous RNA motifs composed of two complementary strands connected by a short linker loop (\textbf{Figure~\ref{fig:structures}c}). Hairpins are often stabilized by highly conserved four-nucleotide loops known as tetraloops, which typically follow the GNRA, UNCG, or CUUG sequence patterns \cite{woese_architecture_1990}, where N denotes any nucleotide and R a purine (A or G). Predicting the native structure of these motifs has traditionally posed a major challenge for MD simulations. The difficulty arises both from sampling limitations---due to the entropic barrier associated with the closing of the loop via Watson-Crick base pairing---and from the highly cooperative nature of the stabilizing interactions formed within the loop. In addition, certain sequences, such as UNCG, remain notoriously difficult to model accurately using current RNA force fields \cite{mlynsky_can_2025}.
These two issues---sampling difficulty and force-field inaccuracy---are tightly linked. Extensive sampling is needed to either fold into the native structure or to reveal its instability, thereby testing the predictive power of the force field.

Several studies have reported extensive simulations of GNRA tetraloops, primarily to demonstrate the capabilities of enhanced sampling techniques \cite{
janecek_w-resp_2021,
zerze_computational_2021,
frohlking_automatic_2022,
mlynsky_toward_2022,
copeland_gaussian_2022,
mlynsky_simple_2023,
gupta_deep_2024,
herron_inferring_2024,
frohlking_deep_2024}. Loops of this family typically fold correctly with AMBER-based force fields, although this observation may depend on the details of the simulation setup. 
By contrast, the more challenging UNCG tetraloops
have frequently been used to benchmark and refine RNA potentials \cite{janecek_w-resp_2021,mlynsky_toward_2022,frohlking_automatic_2022,copeland_gaussian_2022,mlynsky_can_2025}. 
 A study by Bottaro et al.~\cite{bottaro_integrating_2020} integrated NMR data and enhanced sampling simulations to reconstruct a heterogeneous conformational ensemble of the UUCG tetraloop, reconciling experimental observations that were difficult to interpret within a single structural model.
The less common CUUG tetraloop has been examined in fewer studies \cite{copeland_gaussian_2022,oxenfarth_integrated_2023}.
Copeland et al.~\cite{copeland_gaussian_2022} reported a free-energy landscape where the native structure is not the global minimum, which suggests that it is not stable with the current AMBER force field.
Oxenfarth et al.~\cite{oxenfarth_integrated_2023}, instead, report relative short simulations focused on local dynamics and integration of experimental data.

Beyond tetraloops, longer hairpin loops have also been investigated. Rei{\ss}er et al.~\cite{reiser_conformational_2020} studied a hairpin proposed to contain the functional domain of a non-coding RNA. In their work, local conformational fluctuations were explored without attempting global folding. Enhanced sampling was performed using replica-exchange with collective-variable tempering, and maximum entropy integration of NMR data was applied both on-the-fly and in post-processing, enabling reconstruction of ensembles consistent with experiment, including apparently incompatible restraints.
Finally, Bottaro et al.~\cite{bottaro_conformational_2021} used replica-exchange simulations combined with secondary structure restraints to generate conformational ensembles for several stem-loops in the 5$^\prime$ untranslated region of SARS-CoV-2.

\subsection{Internal loops}

Internal loops have been studied using a combination of NMR experiments and MD simulations. In particular, Akinyemi et al.~\cite{akinyemi_nmr_2025} characterized an internal loop flanked by different closing base pairs. Their NMR data enabled the quantification of populations for multiple conformational states. While some MD simulations became trapped in metastable states---highlighting the need for enhanced sampling techniques---other simulations sampled transitions between states but failed to reproduce the experimental populations. These discrepancies suggest that such systems and reference datasets could provide valuable benchmarks for assessing the accuracy of RNA force fields in capturing local dynamics.

Another internal loop was investigated by Taghavi et al.~\cite{taghavi_conformational_2023} using long temperature replica exchange MD simulations (tens of microseconds per replica). The molecule did not unfold during the simulation, although it could have been expected at the highest temperatures. Several force fields were evaluated, and it was concluded that only one of them
could properly model strand slippage, in this case corresponding to 
two alternative non-canonical pairing patterns, in a manner consistent with NMR data.

\subsection{More complex architectures}

Simulations of complex RNA architectures often rely on local sampling initialized from experimentally determined structures, rather than providing exhaustive explorations of the free-energy landscape,
with some notable exceptions.
Enhanced sampling techniques were employed to estimate the folding free energy of an RNA quadruplex (\textbf{Figure~\ref{fig:structures}d}) \cite{pokorna_rna_2025-1}. Using a protocol analogous to those applied in tetraloop folding studies (see Section \ref{sec:hairpin-loops}), they found that sampling was insufficient to fully converge the folding process due to the landscape's complexity. Their results suggest that the expected native structure is not the global free-energy minimum with the AMBER force field.
An RNA pseudoknot was used as a benchmark for
the multithermal-multiumbrella on-the-fly probability enhanced sampling method
\cite{malekzadeh_optimizing_2025}.
Lazzeri et al.~\cite{lazzeri_rna_2023} investigated folding events in RNAs ranging from 20 to 47 nucleotides using ratchet-and-pawl MD, a method that applies a directional bias to drive folding. They observed that the free-energy landscape of RNA is significantly more frustrated than that of typical protein systems. This finding highlights the limitations of directly applying protein-folding simulation strategies to RNA systems.
Other  studies have employed variants of replica exchange to characterize local fluctuations in more complex RNA motifs, such as ribozymes \cite{forget_simulation-guided_2024} and aptamers \cite{autiero_enhanced_2023}.
Finally, the transition pathways between two alternative conformations for a SARS-CoV-2 frameshifting element
were characterized using a combination of transition path sampling and free-energy calculations \cite{yan_heterogeneous_2025}.

\begin{figure}
    \centering
    \includegraphics[width=\linewidth,,trim=0 10 0 0,clip]{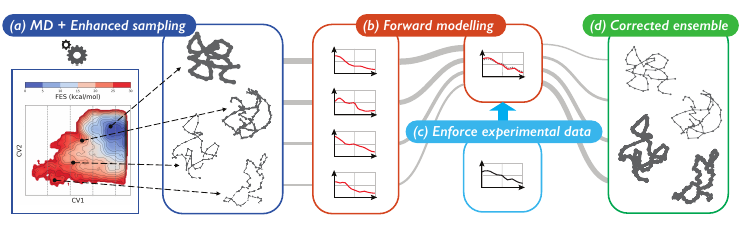}
    \caption{Conformational sampling and integrative approaches for RNA dynamics. \textit{(a)} Molecular dynamics simulations, possibly combined with enhanced sampling, are used to explore the free-energy landscape of an RNA system. This yields a conformational ensemble with associated populations. \textit{(b)} Experimental observables are computed for each structure using forward models. \textit{(c)} Averages over the ensemble are compared with experimental data, which can then be used to reweight the population estimates \textit{(d)}.
Adapted from Bernetti et al.~\cite{bernetti_reweighting_2021}.}
    \label{fig:sampling-integrative}
\end{figure}

A complementary strategy to improve simulation accuracy involves integrating experimental data
(\textbf{Figure~\ref{fig:sampling-integrative}}). SAXS and WAXS data have been particularly valuable in this context. These techniques provide structural information averaged over solution ensembles,
allowing them to probe solution dynamics at low resolution.
Due to their limited information content, SAXS/WAXS profiles are well suited for combination with MD simulations, which supply atomistic detail and help interpret experimental data.
For example, SAXS measurements in the presence and absence of \ce{Mg^2+} were used to infer the relative populations of extended and compact states in a GTPase-associated center \cite{bernetti_reweighting_2021}, and WAXS data were employed to bias the conformational sampling of RNA triplexes \cite{chen_insights_2022}. In the case of a SARS-CoV-2 pseudoknot, SAXS-based selection was used to identify conformations from simulation snapshots that best matched experimental scattering profiles \cite{he_atomistic_2023}. While this selection procedure may suppress conformational diversity---potentially discarding low-populated yet physically relevant structures---it can be effective in avoiding unphysical conformations arising from force-field limitations.
SAXS has also been used to compare force fields in modeling helix-junction-helix motifs \cite{he_refining_2022}, further illustrating its utility in evaluating simulation accuracy for complex RNA topologies.
Integrative modeling approaches have also been applied to characterize the dynamics of group II introns using cryogenic electron microscopy (cryo-EM) data \cite{posani_ensemble_2025}. The authors argued that a significant fraction of RNA structures derived from cryo-EM may contain incorrect base pairing, largely due to the absence of dynamics in the atomistic model. To address this issue, they proposed ensemble refinement as a preferred strategy to generate conformations that are consistent with both experimental data and the known biochemical behavior of RNA.

\subsection{RNA modifications}
\label{sec:modifications}

An increasing number of studies have incorporated non-standard nucleotides into simulations, reflecting the growing recognition of RNA modifications as key regulators of structure and function. Recent specialized reviews provide in-depth overviews of MD applications to study modified nucleotides \cite{desposito_challenges_2022,piomponi_molecular_2023}.
From a technical standpoint, modular force-field strategies---allowing for independent parameterization of modified nucleobases, sugars, or phosphate groups---can greatly simplify the construction of modified RNA topologies. Such modular approaches have been proposed and tested in recent works \cite{love_modxna_2024,silva_characterizing_2025}.

Several studies have combined simulations and experiments to understand or design modified RNA systems. For instance, MD was used to guide the design of antisense oligomers with hydrocarbon-bridged backbones aimed at reducing structural fluctuations and enhancing hybridization with target RNAs \cite{rajasekaran_backbone_2022}. Phosphorothioate substitutions, which replace a non-bridging phosphate oxygen with sulfur, were investigated using a combination of MD and QM/MM simulations \cite{zhang_phosphorothioate_2021}; accurate parameterization of the sulfur atom posed a particular challenge.
Other studies focused on common post-transcriptional modifications. The effect of N6-methyladenosine (\ce{m^6A}) on RNA dynamics was examined using both an energy landscape framework with implicit solvent \cite{roder_structural_2020} and explicit-solvent alchemical approaches \cite{piomponi_molecular_2022}. In the latter, partial charges were refined directly from experimental data, exemplifying a promising strategy for
fine-tuning force-field parameters on thermodynamic data.
Importantly, a later work from the same authors
\cite{piomponi_molecular_2024} showed how
this force-field refinement approach should be based on a set of structural contexts that is heterogeneous enough to be representative of all the interactions that the modified nucleobase can form.
It is also important to include artificial modifications introduced during structural biology experiments. For example, methylations are sometimes added to inhibit catalysis in ribozymes; such chemical modifications should be retained in simulations to faithfully reflect the experimental system and interpret it in a critical manner \cite{forget_critical_2025}.
An integrative approach combining MD, SAXS, and NMR was used to investigate multiple consecutive adenosine-to-inosine (A-to-I) modifications in an RNA duplex \cite{muller-hermes_unique_2025}. This work demonstrated that these edits substantially increase structural dynamics. It also raised concerns about the reliability of inosine parameters in the AMBER force field, which may not suffice for predictive modeling in the absence of integrated experimental data.
Finally, protonation states of nucleobases can influence RNA conformations and base pairing. Some of the studies discussed above employed protonated nucleotides, such as those in References~\citenum{forget_simulation-guided_2024} and \citenum{forget_critical_2025}. Protonation can be conformation-dependent, making constant-pH simulation protocols necessary (Section \ref{sec:constant-ph}). Additional studies of RNA modifications that are relevant for RNA--protein interactions are discussed in Section \ref{sec:protein-rna}.

It is important to keep in mind that force fields for modified nucleotides are usually much less tested than the corresponding force fields for the most common nucleotides. As a consequence, one
should always cross check that the employed parameters have been already tested on a system similar to the one under investigation or do the proper benchmarking against experimental data.

\section{RNA MOLECULES ARE NOT ALONE}

\begin{figure}
    \centering
    \includegraphics[width=\linewidth,trim=0 5 0 5,clip]{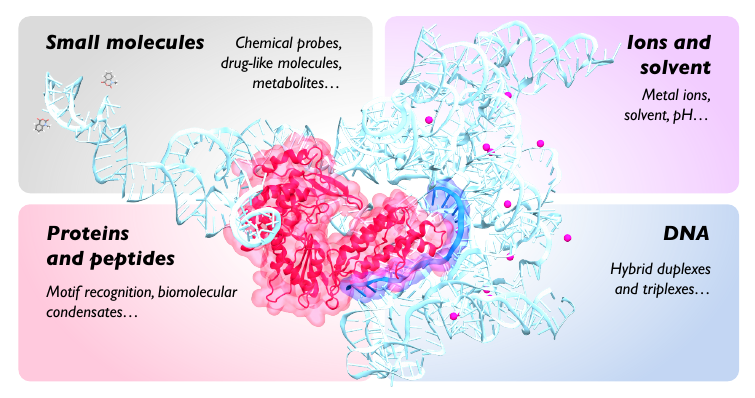}
    \caption{RNA molecules can interact with multiple chemical entities that modulate their function and dynamics. For example, a group II intron prior to reverse splicing (PDB: 6ME0, Reference~\citenum{haack_cryo-em_2019}) forms a complex with both its target DNA (blue) and a maturase protein (red). Simulations should account for the importance of ions (pink) and, depending on the context, small organic molecules (gray).}
    \label{fig:notalone}
\end{figure}

In this section we review recent MD studies of the interaction between RNA and other molecular entities
(see \textbf{Figure~\ref{fig:notalone}}).
In vivo, RNA molecules function within complex environments, where they interact with ions, small molecules, proteins, and membranes.
Most of the studies discussed here investigate how binding events influence RNA structure and dynamics, thereby revealing mechanisms of conformational adaptation and stabilization that are often inaccessible to experimental techniques alone.
The accuracy of these simulations strongly depends on the quality of the underlying force fields.
In particular, subtle inaccuracies in describing electrostatics can have amplified effects, and in some cases polarization has been shown to be necessary to reproduce experimental observations.
Nevertheless, when experimental data are available and agreement can be achieved, MD simulations provide unique insights into the conformational transitions and coupling mechanisms underlying RNA--partner interactions, complementing and extending structural information from experiments.

\subsection{Interaction of RNA with ions}

RNA molecules are polyanionic in physiological conditions and therefore tend to interact strongly with cations. MD simulations in explicit solvent are typically performed under net-neutral conditions, with a compensating background of ions. RNA--ion interactions have been understood through two paradigms: as a diffuse electrostatic cloud, termed ion atmosphere, acting as an overall screening environment, or as site-specific interactions that can be direct (inner-sphere) or solvent-mediated (outer-sphere). While ion atmosphere effects can be approximated implicitly using mean-field theories, site-specific interactions require explicit modeling and a special attention to the force-field parameterization.

The most relevant monoatomic ions that interact with RNA are \ce{Mg^2+}, \ce{K+}, \ce{Na+} and \ce{Cl-}, although other species may be abundant in experimental buffers \cite{leontis_ions_2012}. Anions mostly interact indirectly through participation in the electric double layer that forms around the negatively charged RNA.

Ab initio simulations are useful for determining geometric features of RNA--ion interactions \cite{kolev_sodium_2022}. These can help devising stereochemical criteria and exclusion zones for the critical assessment of ion placement in crystallographic and cryo-EM data, as well as refining classical force-field parameters. Site-specific binding to cations is more likely to occur with non-bridging backbone oxygen atoms, carbonyl oxygen and basic nitrogen atoms of the nucleobase (in particular N7 and O6 atoms on guanine, O4 on thymine and uracil, and O2 on cytosine) and to a lesser extent ribose oxygen \cite{cruz-leon_rna_2022}. The backbone follows a direct Hofmeister series, with inner-sphere binding affinity growing with cation charge density (\ce{K+}~<~\ce{Na+}~<~\ce{Mg^2+}) while the nucleobase sites follow a reversed Hofmeister series (\ce{K+}~>~\ce{Na+}~>~\ce{Mg^2+}) \cite{cruz-leon_hofmeister_2020}.

Simulations have confirmed the role of both mono- and divalent cations in modulating RNA catalytic mechanisms, through conformational stabilization, direct stabilization of the transition state, or by affecting the $\mathrm{p}K_\mathrm{a}$  of nearby groups \cite{ekesan_rna_2022, ahsan_emerging_2025}.

\subsubsection{Interaction with monovalent cations}

While the most abundant intracellular monovalent cation is \ce{K+}, \ce{Na+} is often preferred in MD studies to match the experimental buffer in which the structure of interest was resolved.
Yet, site selectivity among monovalent ions has been shown to influence RNA unfolding in steered MD simulations: the smaller \ce{Na+} ion exerts a stronger stabilizing effect than \ce{K+}, likely due to its preferential interaction with the phosphate backbone rather than the major groove \cite{henning-knechtel_differences_2022}.
This selectivity also explains the usual preference of G-quadruplexes for potassium, although their stability in silico is greatly affected by the precise ion parameterization,
and explicit polarizability may be needed to fully understand the underlying mechanisms \cite{lemkul_same_2020,poleto_differences_2023}. When compared to the prevalent B-form DNA, RNA A-form duplexes display a higher affinity for ions and distinct binding patterns, notably a zipper-like closing of the major groove \cite{cruz-leon_rna_2022}.

Initial placement of monovalent ions is usually not a concern due to their fast dynamics and short residence times (ps--ns) \cite{cruz-leon_hofmeister_2020} when compared to modern simulation time scales and RNA conformational changes. Extensive MD showed a moderate effect of cationic species and concentration on the compactness of A-form RNA duplexes, the critical aspect remaining the RNA force field itself \cite{kuhrova_sensitivity_2023}.

Ionic conditions are also expected to affect RNA interaction with other molecules. For instance, a recent work has demonstrated the key importance of monovalent ions in modulating the affinity of a poly-pyrimidine recognition protein domain \cite{rozza_monovalent_2023}. Notably, using excess \ce{NaCl}---better matching physiological conditions, instead of merely compensating the solute charge---impacted the binding affinity in a sequence-dependent manner.

\subsubsection{Interaction with divalent cations}

Due to their high charge density, divalent cations such as \ce{Mg^2+} or \ce{Ca^2+} pose significant challenges for atomistic simulations.
Because of polarization effects, it is unlikely that classical, point-charge additive force fields can be tuned to describe divalent ions with sufficient accuracy \cite{liebl_development_2023}. Nevertheless, improvements can be made within the limited classical force fields, and mitigate the over-binding artifacts often observed for cations parameterized in bulk conditions. Possible strategies include electronic-continuum-correction charge rescaling \cite{puyo-fourtine_amber-ol15-ecc_2025}, modified Lennard-Jones combination rules for ion-RNA atom pairs \cite{grotz_optimized_2021,grotz_magnesium_2022}, more expressive LJ potentials with an additional $r^{-4}$ attractive term \cite{koca_findik_binding_2024} or multi-site models \cite{peng_developing_2021}.

Among metal cations, \ce{Mg^2+} is an outlier in terms of ligand exchange kinetics, with first-shell water residence on the microsecond time scale and transition from outer- to inner-sphere coordination from the micro- to the millisecond time scales. It is thus extremely challenging to obtain the equilibrium distribution of inner-sphere bound \ce{Mg^2+} on biomolecules and such goal is often precluded \cite{
cruz-leon_hofmeister_2020,mainan_magnesium_2024,puyo-fourtine_amber-ol15-ecc_2025} unless one uses targeted enhanced sampling methods such as biased MD \cite{languin-cattoen_predicting_2024} or grand canonical Monte Carlo \cite{kognole_contributions_2020,sengul_influence_2023,baral_grand_2024}. Alternatively, a recent hybrid strategy termed dynamic counterion-condensation was used to model outer-sphere \ce{Mg^2+} explicitly, while treating solvent and monovalent ions implicitly \cite{ramachandran_dynamic_2022,mainan_magnesium_2024}. Inner-sphere \ce{Mg^2+} ions were modeled with ad hoc contact-based potential energy terms, hence requiring prior knowledge of the native structure. Despite this limitation, the authors were able to support the crucial role of \ce{Mg^2+} in stabilizing intermediates on RNA folding pathways \cite{mainan_magnesium_2024} and modulating native fold breathing dynamics \cite{ramachandran_dynamic_2022}. 

Simulations combining transition path sampling with deep neural networks elucidated the binding pathway to the phosphate backbone \cite{neumann_artificial_2022}. By broadening the Lennard-Jones parameter search and customizing RNA--ion rules, Grotz and coworkers have developed a model, microMg, that reproduces kinetic properties alongside the usual thermodynamic and geometric features \cite{grotz_optimized_2021}. Conveniently, they also found a set of parameters, nanoMg, whose faster shell-exchange dynamics in the nanosecond range can be leveraged.

\subsubsection{Constant-pH simulations}

\label{sec:constant-ph}

As mentioned in Section~\ref{sec:modifications}, protonated nucleotides can influence RNA structure. In experiments, where the pH is typically controlled, the protonation state of nucleotides may depend on the local conformation. Performing constant-pH MD simulations, however, remains challenging. A recent RNA-focused example is presented in Reference~\citenum{silva_characterizing_2025}, where parameters for protonated nucleobases were developed and an implicit solvent-based scheme was tested, demonstrating that $\mathrm{p}K_\mathrm{a}$ shifts induced by RNA chain length can be reproduced.
In practice, it is important to note that protonated or deprotonated nucleobases are often not annotated in experimental structures. Constant-pH methods can help identify such protonation states in experimental data and guide further simulations, either with fixed protonation (standard MD) or under dynamic titration conditions (constant-pH MD).
Importantly, protonated nucleobases can modulate and potentially compete with cation binding. This suggests that explicitly accounting for protonation states may be essential for accurately characterizing ion binding sites in RNA.

\subsection{Interaction of RNA with small molecules}

RNA structure and function can be modulated by the interaction with small molecules. We here cover the case of 
endogenous metabolites and drugs, as well as reagents used in chemical probing experiments.

\subsubsection{Riboswitches and metabolites}
\label{sec:riboswitches}

Riboswitches are structural motifs located in the untranslated regions of mRNAs  
that bind a cognate ligand with high specificity and consequently modulate gene expression
by controlling transcription, translation, or alternative splicing.  
The preQ\textsubscript{1} riboswitch, one of the smallest known, has been used as a benchmark  
for enhanced sampling methods to accelerate RNA--ligand binding \cite{serra_path-based_2025}.  
In a mixed experimental and computational study, plain MD simulations were employed  
to characterize the stability of the region controlling gene regulation \cite{schroeder_riboswitch_2023}.  
The same riboswitch was also studied using artificial intelligence-based methods  
to characterize dissociation pathways of both cognate and synthetic ligands,  
including the effect of distal mutations (located away from the binding site) on their affinity \cite{wang_interrogating_2022}.  
A similar approach was used to characterize a series of ligands binding the ZTP riboswitch \cite{fullenkamp_machine_2025}.  
The effect of distal mutations on the folding pathway of a riboswitch was also investigated in Reference~\citenum{chyzy_molecular_2024},  
where a combination of umbrella sampling and replica exchange with solute tempering was used  
to simulate the binding of neomycin to an artificial riboswitch with various mutations.  
Interestingly, this work discussed the effect of a possibly protonated nucleobase
(see Sections \ref{sec:modifications} and \ref{sec:constant-ph}).

A series of studies investigated the binding of guanidine riboswitches  
with their cognate ligand.  
Two of these papers focused on the guanidinium-II riboswitch, which binds  
its ligand through a conserved ACGA loop, first showing that  
ligand binding leads to a collapse in the conformational flexibility of the loop \cite{steuer_guanidine-ii_2021},  
and later discussing the design, simulation, and experimental validation of bivalent ligands  
that can simultaneously bind two copies of the same aptamer domain, stabilizing the intermolecular complex \cite{steuer_cooperative_2024}.
In another study, the guanidinium-I riboswitch was analyzed to investigate the interplay  
between \ce{K+} and guanidinium binding, showing that, in this system,  
the presence of \ce{K+} ions in the proper configuration is required  
to stabilize the bound pose of the ligand \cite{franke_atomistic_2024}.  
Yet another guanidinium-binding riboswitch was characterized with respect to its binding specificity  
in Reference~\citenum{negi_molecular_2021}. Ligand specificity was also studied in Reference~\citenum{kumar_mechanism_2023},  
where alchemical calculations were used to assess the impact of mutations in the riboswitch sequence.

Purine riboswitches have been studied in multiple works.  
One purine riboswitch has been used as a benchmark system for force-field modifications \cite{chen_rna-specific_2022}.  
A different purine riboswitch was investigated to characterize the impact of the ligand and \ce{Mg^2+}  
on conformational dynamics \cite{chaudhury_conformational_2025}.  
Whereas no large conformational changes are expected on the reported time scales (500~ns),  
the observed dynamics was found to be consistent with chemical probing experiments.  
The interplay between ligand and \ce{Mg^2+} binding was also explored in Reference~\citenum{bao_potential_2022}.

The SAM riboswitch has been characterized using either Gaussian accelerated MD \cite{chen_decoding_2022},  
or a meta\-dynamics-\allowbreak based approach aimed at accelerating the dynamics of tertiary contacts \cite{prajapati_exploring_2022}.  
In the latter study, free-energy landscapes in the presence and absence of the ligand,  
and under different ionic conditions, were found to be consistent with SAXS experiments.
The flavin mononucleotide riboswitch was used as a benchmark system to compare ligand binding to a riboswitch versus a protein, highlighting both similarities and differences in RNA and protein recognition \cite{bosio_similarities_2024}. Based on this comparison, the authors suggested that while electrostatic contributions are comparable, RNA flexibility plays a more prominent role than in the protein target.

Finally, MD simulations were used to construct structural models of co-transcriptional folding  
intermediates, based on secondary structures obtained from chemical probing data \cite{cheng_cotranscriptional_2022,hertz_effect_2024},  
demonstrating how MD simulations can provide a microscopic interpretation of very coarse-grained experimental measurements.

\subsubsection{Drug-like molecules}

RNA molecules are increasingly considered potential drug targets, spurring interest in designing drug-like compounds that bind RNA structural motifs with high affinity and specificity. Two fragment-based approaches have recently been introduced---one combining MD and Monte Carlo simulations \cite{kognole_silcs-rna_2022}, the other employing enhanced sampling techniques \cite{panei_identifying_2024}---and tested on heterogeneous sets of RNA motifs.
The binding of drug-like molecules to CUG repeats has been studied using a combination of docking tools and umbrella sampling simulations in explicit solvent \cite{wang_dynamic_2023}.

A general concern that applies to all such studies---including those targeting the binding of small molecules to riboswitches (see Section \ref{sec:riboswitches})---is the accuracy of classical force fields. While force fields for standard nucleotides are relatively well tested, those for small molecules are often less rigorously validated. This discrepancy can bias the comparison of binding affinities across different compounds, complicating rational drug design.
A recent example studying the interaction of a drug-like molecule with CAG repeats illustrates how polarization effects can significantly influence the predicted binding pose, underscoring the challenges of relying solely on classical force fields for these systems \cite{hoang_refining_2025}.

\subsubsection{Interaction with chemical probes}

Chemical probing experiments use small reactive molecules to modify RNA, providing indirect insights into its structure. Given the widespread use of these techniques, several groups have attempted to gain a microscopic understanding of how 
the employed reagents interact with specific RNA motifs, using MD simulations to model physical binding \cite{mlynsky_molecular_2018,hurst_sieving_2021}.
Simulations including multiple copies of the reagent---thus allowing for potential cooperative effects---have shown that physical binding alone, at concentrations typical of probing experiments, may significantly alter RNA dynamics \cite{calonaci_molecular_2023}. More recently, a group has taken on the challenging task of modeling the covalent binding of a SHAPE reagent using QM/MM simulations \cite{hognon_molecular_2025}.

\subsection{RNA molecules in biomolecular condensates}

Biomolecular condensates organize the cellular environment into membraneless compartments containing very high concentrations of biopolymers.
MD simulations of RNA oligomers in water, or in the presence of increasing concentrations of oligopeptides, have probed the microscopic determinants of RNA--peptide interactions.
They revealed that arginine aggregates with RNA more than lysine due to its ability to form multivalent interactions \cite{paloni_arginine_2021}.
Enhanced sampling simulations, using a standard protocol for the reversible folding of RNA stem-loops (see Section~\ref{sec:hairpin-loops}), have shown that peptides can significantly influence the stability of both base-paired regions and structured loops \cite{boccalini_exploring_2025}.
Simulations of larger-scale condensates---comprising up to hundreds of full-length proteins and dozens of RNA polymers---have primarily been used to benchmark protein, RNA, and water force fields \cite{sarthak_benchmarking_2023}.

\subsection{RNA--protein complexes}
\label{sec:protein-rna}

Several groups have studied the interaction of structured proteins with short single-stranded RNAs. In particular, the binding of an RNA recognition motif to single-stranded RNA has been investigated using long, unbiased molecular dynamics simulations \cite{krepl_spontaneous_2022}. The authors found that overly strong stacking interactions in the AMBER force field can overstabilize the unbound RNA, thereby preventing spontaneous binding. To observe binding events, they introduced system-specific modifications to weaken stacking interactions.
A related line of research has focused on how N6-methylation of adenosine (see Section~\ref{sec:modifications}) affects recognition by \ce{m^6A} reader proteins \cite{krepl_recognition_2021,li_atomistic_2021,zhou_specific_2022,
piomponi_molecular_2024}. One recent study also employed QM/MM simulations to characterize the enzymatic mechanism of methylation itself \cite{corbeski_catalytic_2024}.

Docking simulations have explored RNA--aptamer interactions with structured proteins. In one case, the aptamer structure was built de novo, and the resulting complex was characterized with standard MD simulations \cite{diakogiannaki_targeting_2025}. The designed aptamer included a modified nucleotide (see Section~\ref{sec:modifications}), which was not explicitly modeled. In another study, supervised MD simulations were employed to perform docking in an explicit solvent environment \cite{pavan_investigating_2022}.

Other studies have focused on cases where protein structure is strongly influenced by RNA binding. For example, the interaction between a boxB loop and a peptide was used as a model system to compare water models and restraining protocols in MD simulations \cite{vollmers_advanced_2024}. The complex formed by the HIV trans-activation response element and a cyclic peptide optimized for binding has also served as a benchmark for developing machine learning-based collective variables \cite{kumari_enhanced_2025}.

\subsection{RNA--membrane interactions}

The interaction of RNA molecules with membranes is an emerging field of research.
There are two main cases of application. First, ionizable lipids are usually employed
to deliver RNA molecules in gene therapy and RNA vaccines.
Their interaction with negatively charged nucleic acids can be modulated
by pH.
A few MD works have characterized the association of RNA structures with those membranes
\cite{rissanou_complexation_2020,paloncyova_role_2021}.
Recently, it has been suggested that endogenous RNA molecules
can be transported to the extracellular surface of living cells.
MD simulations have then be used to characterize the physico-chemical determinants
of RNA--membrane interactions, trying to elucidate its relationship with RNA
sequence and structure
\cite{di_marco_all-atom_2025,singh_rna_2025}.

\subsection{Catalysis}

A number of studies used MD simulations to rationalize catalysis in ribozymes.
These studies normally used a combination of classical MD and QM/MM simulations.
For reasons of space, we  
refer the reader to other reviews that cover this topic \cite{borisek_establishing_2023,ahsan_emerging_2025}.

\section{DISCUSSION}

The Sections above provided a general overview of the field, summarizing a number
of recent publications. In this Section, we highlight emerging techniques
and speculate on the future of atomistic MD simulations applied to RNA systems.

\subsection{Emerging artificial-intelligence and machine-learning techniques}

\begin{figure}
    \centering
    \includegraphics[width=\linewidth,trim=0 0 0 0,clip]{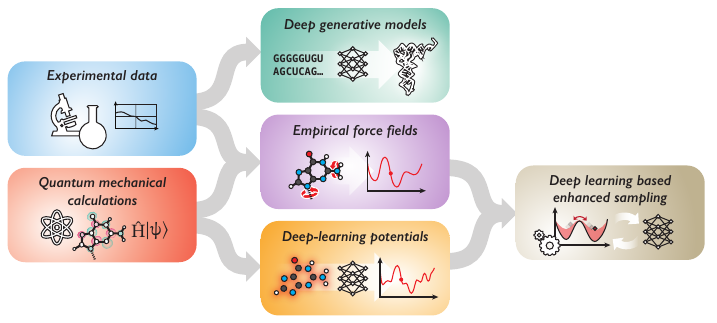}
    \caption{
    Deep-learning methods for RNA structure and dynamics. Deep-learning generative methods are typically trained exclusively on experimental data, and deep-learning potentials are typically trained on quantum mechanical calculations. Their applicability to RNA is still in its infancy, due to the lack of training data and to the high computational cost, respectively. Simulations with empirical force fields blend both empirical and quantum-mechanical information. Deep-learning methods can also be used to effectively accelerate sampling in molecular simulations.
    }
    \label{fig:artificial-intelligence}
\end{figure}

Deep learning-based methods are emerging as tools to study conformational dynamics of RNA systems
(see \textbf{Figure~\ref{fig:artificial-intelligence}}). Below we dissect the different ways in which deep-learning methods can assist RNA modeling.

\subsubsection{Bottom-up artificial intelligence}

Machine learning offers a route to bridge costly ab initio calculations with the computationally efficient traditional MD \cite{unke_machine_2021}. The promise is to train a highly accurate potential energy function from quantum-mechanical calculations that can be generated as needed, then scale to larger and longer simulations while avoiding the daunting task of solving Schrödinger's equation. Classical force fields---including polarizable ones---are limited by their functional form, and may not be expressive enough to cover a wide range of RNA systems \cite{mlynsky_can_2025}. Modern artificial-intelligence (AI) paradigms including deep neural networks (NNs) as universal approximators can in principle capture the complex dependence of the ground state energy to atomic coordinates. However, all-NN forcefields have limited generalizability and transferability to larger or chemically diverse systems, and they may ignore tried-and-true physical insights. Hence, physics-based functional forms remain actively pursued, either alone \cite{spoel_evolutionary_2025} or in hybrid schemes with NNs \cite{illarionov_combining_2023}.
Comparable strategies are used to parameterize CG models from all-atom data.

\subsubsection{Top-down artificial intelligence}

Although bottom-up approaches offer interesting prospects, the most significant successes of deep learning in structural biology have come from top-down approaches, where the models are directly trained on experimental data. Following the success of AlphaFold 2 for protein native structure prediction, several deep learning-based models have been proposed for RNA structure prediction, with accuracies still lagging behind proteins due do the relative scarcity of the training data \cite{schneider_when_2023}.
A recent review covers these methods~\cite{wang_deep_2025},
and community benchmarks confirm that human expertise remains essential for blind RNA-structure prediction \cite{bu_rna-puzzles_2025}.
An alternative top-down machine learning approach takes advantage of the vast amounts of unannotated sequencing data to pre-train general-purpose representations of RNA sequences in a self-supervised way. These so-called foundation models can be used in many downstream tasks after fine-tuning~\cite{wang_deep_2025}. Their capability to capture RNA structure, however, remains unproven~\cite{papazoglou_predicting_2025}.

Far less attention has been given to RNA dynamics. Generative frameworks, such as diffusion or flow matching, can in principle produce conformational ensembles,
but remain limited by the available training data. A recent approach \cite{tarafder_rnabpflow_2025} conditions the generative process on a pool of prescribed secondary structures, which are easy to generate, then scores and refines the resulting ensemble. The results were validated on ensemble predictions using a separate CG method \cite{li_rnajp_2023}. To our knowledge, no standardized, experimentally validated dataset is available for RNA dynamics and one must rely on low-dimensional, system-specific experimental evidence through integrative approaches (NMR, SAXS, chemical probing, etc.). In this context, MD simulations, especially when combined with
experiments, may offer a substantial and reliable source of training data for generative models.

\subsubsection{ML-assisted enhanced sampling}

Another promising direction is the integration of traditional simulations with AI-based enhanced sampling methods. %
The goal is to improve sampling efficiency, hence precision, enabling faster generation of structures from the Boltzmann distribution defined by the chosen force field. Several studies discussed in this review have explored these approaches to investigate the folding of small RNA motifs \cite{herron_inferring_2024,frohlking_deep_2024,gupta_deep_2024}, RNA--ion interactions \cite{neumann_artificial_2022}, RNA--ligand \cite{wang_interrogating_2022,fullenkamp_machine_2025} and RNA--protein \cite{kumari_enhanced_2025} binding.

\subsection{The future of MD simulations of RNA}

Molecular dynamics (MD) simulations based on empirical force fields have been used for decades to study RNA dynamics. Their ability to reproduce experimental data has fluctuated over time and remains highly system-dependent. This variability can be discouraging---particularly for newcomers---since substantial expertise is often required to judge whether a given RNA system can be reliably modeled with current force fields. One of the goals of this review is to help mitigate this challenge by surveying recent developments and outlining the strengths and limitations of current approaches.
Given the intrinsic time-scale limitations of standard MD, enhanced sampling methods remain central. Techniques such as replica exchange, metadynamics, umbrella sampling, and AI-assisted approaches are increasingly employed to capture rare conformational transitions in RNA folding, ligand binding, and catalysis---processes  often inaccessible through conventional simulations. While the technical details of these methods lie beyond the scope of this review, recent successful applications are discussed throughout.

The predictive power of atomistic simulations has steadily improved, driven by systematic validation against experiments and refinement of force-field parameters. Modern RNA force fields now incorporate insights from quantum mechanical calculations alongside empirical tuning informed by experimental observables.
Still, some RNA  classes---such as G-quadruplexes, UNCG tetraloops, and kink-turn motifs---remain challenging, with no universal correction yet identified. This suggests that some limitations may be intrinsic to current empirical frameworks. Although the parameter space of classical force fields is vast and only partially explored,  approximations such as fixed charges and non-polarizable models
imply that their performance may be nearing a plateau. Future advances could come from the large-scale reparameterization of polarizable force fields, which better capture electronic response, or from machine learning-based approaches. Neural network potentials, in particular, offer flexible functional forms and can be retrained across diverse chemical and structural environments,
at the price of disregarding physical wisdom and requiring larger training sets.
To address persistent limitations, the field increasingly relies on integrative modeling strategies combining experimental restraints with simulation. These approaches not only mitigate force-field inaccuracies but also provide atomistic interpretation of sparse or low-resolution data. Importantly, they are model-agnostic and can be applied to classical, polarizable, or machine-learned potentials alike.

Finally, simulated conformational ensembles---especially when refined using experimental data---are emerging as valuable training inputs for generative AI. %
Continued progress will depend on close collaboration between computational and experimental communities. Shared benchmarks, open datasets, and reproducible workflows \cite{amaro_need_2025} will be essential to assess emerging methods and translate simulation insights into testable hypotheses.

\section*{FUTURE ISSUES}
\begin{enumerate}
\item Improved tools are needed to prepare MD simulations of RNA systems starting from experimental structures. These should support, among other tasks: enforcing correct biochemical features (e.g., base pairing), semi-automated identification of protonated and modified nucleotides, and automated analysis of divalent cation coordination.

\item Future RNA force fields should leverage advances in machine learning and be trained on a combination of quantum mechanical calculations and experimental data.

\item Curated databases of experimental measurements and molecular dynamics trajectories will be essential to advance the field.

\item The functional form of RNA force fields may need to be extended to include polarizability or replaced by general-purpose deep learning potentials.

\item Accurate MD simulations will play a key role in supporting experimental techniques that yield sparse or ensemble-averaged information, particularly in the context of integrative modeling.

\item Enhanced sampling methods must become more automated and accessible in order to efficiently capture slow or rare RNA conformational transitions.

\item In a field where high-resolution reference structures are scarce, MD simulations may serve as a valuable source of training data for generative models of RNA structure---or, more ambitiously, RNA dynamics.

\end{enumerate}

\section*{DISCLOSURE STATEMENT}
The authors are not aware of any affiliations, memberships, funding, or financial holdings that
might be perceived as affecting the objectivity of this review. 

\section*{ACKNOWLEDGMENTS}
O.~L.-C. acknowledges the European Union’s Horizon 2023 research and innovation programme under the Marie Skłodowska-Curie grant agreement No.~101152924.

\bibliographystyle{unsrt-initials} %
\bibliography{references_clean}
\end{document}